# Direct observation of near-field induced resonant electron tunneling in a sub-nanometer plasmonic gap

Shuyi Liu[1], Martin Wolf[1], Takashi Kumagai[1,2]*

[1]*Department of Physical Chemistry, Fritz-Haber Institute of the Max-Planck Society, Faradayweg 4-6, 14195 Berlin, Germany.*

[2]*JST-PRESTO, 4-1-8 Honcho, Kawaguchi, Saitama 332-0012, Japan.*

*Corresponding author: kuma@fhi-berlin.mpg.de

Localized surface plasmon resonance (LSPR) excitation of nanostructures and charge transfer in plasmonic nanocavities plays a central role in nanoscale optoelectronics and in applications for plasmonic devices. However, the direct observation of near-filed induced charge transfer has remained as a challenging experiment. Here we present LSPR-assisted resonant electron tunneling from an Ag or Au tip to the image potential states of a Ag(111) surface using scanning tunneling microscopy (STM). The LSPR-assisted tunneling process results in an incident photon-energy dependent red-shift of the field emission resonances (FERs) in the gap. Using the precise control of the gap distance in the STM junction we demonstrate tuning of the relative contribution from the LSPR-assisted and the normal STM electron tunneling processes. Furthermore, the FER intensity mapping of local defects on the surface allows assigning unambiguously the respective FER levels with and without laser excitation.

Optical excitation of metallic nanoparticles and nanostructures through localized surface plasmon resonance (LSPR) enables many useful applications such as nanoscale nonlinear optics,[1] single-molecule detection,[2] sensitization of photovoltaics,[3] and enhancement of molecular luminescence,[4] photoelectrochemistry,[5] and photocatalysis.[6,7,8,9] LSPR-induced hot carrier generation and electron transfer are of particular importance in plasmonic device applications.[10,11] However, the direct observation of LSPR-induced charge transfer in nanocavities is a challenging experiment and the transfer mechanism across a dielectric or vacuum gap remains poorly understood.

The combination of STM and optical excitation has been used to study optical properties of plasmonic junctions,[12,13,14,15,16,17,18] which allows precise control of the vacuum gap (tunneling barrier) even at sub-nanometer distance as well as high-resolution imaging of nanoscale structures. In addition, LSPR-excitation in STM junctions has also gained increasing attention in nanoscale science and technology for its microspectroscopic applications such as single-molecule luminescence spectroscopy[19,20,21,22,23,24] and tip-enhanced Raman spectroscopy.[25,26,27,28,29] Although the plasmonic excitation in the STM junction can significantly affect the electron tunneling properties, the influence of the strong near-field on electron tunneling in sub-nanometer gaps has not been scarcely investigated and a clear mechanism of LSPR-assisted resonant tunneling remains to be clarified.[30]

In this Letter we report the LSPR-assisted tunneling in an Ag or Au tip–Ag surface junction induced by CW laser excitation. The LSPR-assisted tunneling occurs resonantly from the STM tip into the image potential states of a Ag(111) surface which are observed as multiple peaks in field emission resonance (FER) spectroscopy with STM.[31,32]. This Rydberg-like series of electronic states formed in the Coulomb-like image potential serves as a simple model system to investigate resonant charge transfer in the sub-nanometer plasmonic gap. It is found that the LSPR-assisted tunneling results in pronounced red-shift of the FER peaks depending on the incident photon-energy. Furthermore, the gap distance dependence of the FER spectra under illumination reveals the relative contribution from the LSPR-assisted and the resonant STM tunneling into the image potential states.

The experiments were performed in an ultra-high vacuum chamber (base pressure $<5\times10^{-10}$ mbar) equipped with a low-temperature STM/AFM (modified UNISOKU USM-1400) operated with a Nanonis SPM controller. All measurements were performed at 78 K. A chemically etched Au or Ag tip was used (the Ag tip was supplied by UNISOKU Ltd.). The tips were cleaned by $Ar^+$ sputtering before measurement. The bias voltage was applied to the sample with the tip at ground. The Ag(111) surface was cleaned by repeated cycles of $Ar^+$ sputtering and annealing up to 670 K. The laser beam (446 nm: CW diode laser, 532 nm: CW diode laser, 632 nm: CW HeNe laser) was focused to the STM junction using an *in-situ* Ag-coated parabolic mirror (numerical aperture of ~0.6) mounted on the cold STM stage. The spot diameter on the tip apex was estimated to be about 1 μm. The beam alignment and focusing were performed precisely with piezo motors (Attocube GmbH) attached to the parabolic mirror, which allow three translational and two rotational motions of the parabolic mirror. The incident light was linearly polarized along the tip axis. The FER (d$I$/d$V$) spectra were recorded using a lock-in amplifier with a modulation of 20 mV$_{rms}$ at 983 Hz.

**Figure 1a** shows the FER spectra (the solid curves) of an Ag tip–Ag(111) surface junction with and without 633-nm laser excitation ($hv = 1.96$ eV). The multiple peaks in the spectra correspond to a series of the image potential states of the Ag(111) surface which arise from the many-body screening of an electron in front of the surface by conduction electrons in a metal.[33, 34] We find that the peaks are largely red-shifted under laser illumination and, in particular, the shift of the first peak (~1.9 eV) is nearly identical to the incident photon energy. We denote the FER levels by $n$ and $n'$ (= 1, 2, 3...) with and without laser excitation, respectively. Noted that the d$I$/d$V$ spectra were recorded in the constant current mode which allows to measure a relatively wide voltage window, which causes a continuous vertical displacement (retraction) of the tip (the dashed lines, right axis in **Fig. 1a**). **Figure 1b** shows the incident power dependence of the FER spectra and the power density varies from 0 (top) to 2.67 mWµm$^{-2}$ (bottom). It is clear that the peak around 2.2 (4.2) V grows (diminishes) as the incident laser power increases. The shift of $n = 1$ peak shows a clear correlation with the incident photon energy. As seen in **Fig. 1c**, when the junction is illuminated by a 532-nm laser ($hv = 2.33$ eV), the peak is shifted to a lower energy by ~2.3 eV. In addition, it is shifted by 2.8 eV with 446-nm ($hv = 2.78$ eV) excitation. Furthermore, it is found that an Au tip exhibits a similar red-shift of the FER peaks with excitation at 633 nm (*cf*. **Fig. 4a**), whereas no change occurs with excitation at 532 nm. This observation suggests that the process is mediated by LSPR excitation in the junction as the plasmon of the Au tip cannot be excited with excitation at 532 nm because of the ($5d \rightarrow 6sp$) interband transition.[35] Although the peak voltage and intensity of the FER spectra are slightly affected by tip conditions,[33, 34] the overall behavior discussed above remains essentially unchanged by the tip (*e.g.*, *in-situ* tip preparations or use of different tips).

**Figure 2** shows schematic models of the STM junction and the electron tunneling to the image potential state of the Ag(111) surface with and without laser excitation. FERs in the junction represent lateral quantization of surface state electrons above the vacuum level of the surface.[36, 37] In the absence of laser excitation, the peaks in the FER spectra appear when the bias voltage ($V_s$) of the STM matches the respective resonance level (**Fig. 2a**). As mentioned above, the gap distance ($d$) continuously increases with increase in the bias voltage (see **Fig. 1a**). As the incident power increases, the LSPR-assisted tunneling becomes possible through increment of near-field enhancement in the junction (**Fig. 2b**), resulting in the new (red-shifted) peaks in the FER spectra. Similar photo-assisted electron tunneling into a molecular resonance in an STM junction has also proposed previously.[30] According to this picture, one would expect that the peak caused by the normal FER process (resonant STM tunneling) also appears when the bias voltage matches the first resonance level (as indicated by the dashed arrow in **Fig. 2c**), which is indeed observed at a low power density as seen in **Fig. 1b**. However, this process should become negligible if the LSPR-assisted tunneling dominates. As can be seen in the tip displacement in **Fig. 1a**, the gap distance at the FER voltages under illumination is larger by about 1 nm than that in the absence of illumination. This elongation of the gap distance gives rise to a much wider barrier width at the voltages of resonant STM tunneling, suppressing the electron tunneling into the image potential states from the Fermi level of the tip. On the other hand, the LSPR-assisted tunneling is feasible and becomes even dominant due to the thinner barrier than that of the normal STM tunneling (**Fig. 2c**).

We find that the relative contribution from the normal STM tunneling and the LSPR-assisted process can be tuned by varying the gap distance of the junction. **Figure 3a** shows the FER spectra for an Ag tip with 532-nm excitation recorded at different gap distances (defined by the set current in the figure). As the gap distance decreases, the peaks of higher FER levels ($n \geq 2$) are gradually blue-shifted (*e.g.*, $n = 2$ peak of the LSPR-assisted process which is marked by the black bars in **Fig. 3a**). Additionally, a peak that does not significantly vary with decrease in the gap distance is discernible above the set current of 20 nA (marked by the red bar in **Fig. 3a**). This peak is attributed to the resonant STM tunneling to $n = 1$ image potential state, which competes with the LSPR-assisted process (**Fig. 3b**). A similar gradual blue-shift of the FER peaks can be observed in the absence of illumination[34] and explained by a Stark shift of image potential states caused by the strong electric field in the junction,[38] and the energy spacing increases as the potential becomes steeper in a smaller gap distance (as illustrated by the dashed blue lines in **Fig. 3b**). The Stark shift is relatively weak at $n = 1$ level but the peak width is broadened compared to that at a large gap distance (see the top panel of **Fig. 3a** for the spectrum at a small gap distance without laser excitation) presumably due to hybridization between image potential states and bulk states of the tip and surface.[38] The peak broadening also occurs $n' = 1$ level as the gap distance is reduced.

It is rather obvious that $n = 1$ peak of the FER spectra under illumination arises from the LSPR-assisted tunneling into the first image potential state. In order to assign the other peaks, we measured the STS mapping at the FER peak voltages with and without illumination. **Figure 4a** shows the FER spectra of an Au tip–Ag(111) surface with 633-nm excitation. As mentioned above, the results are similar to the Ag tip. Because each FER is sensitive to a local structure (or defect) of the surface which causes a modulation of the local work function,[34, 39, 40, 41, 42] we mapped the FER intensity over a monoatomic step and an intrinsic defect on Ag(111) (the topographic image of the measurement area is shown in **Fig. 4b**). **Figures 4c–e** displays the FER intensity maps at $n$ = 2, 3, 4 peaks without illumination (labeled c, d, e in **Fig. 4a**), which exhibit unique features at the defect structure. **Figures 4f–h** show the maps at $n'$ = 2, 3, 4 peaks under illumination (labeled f, g, h in **Fig. 4a**). It is clear that each intensity distribution at the same ordinal number of the FER peaks shows very similar features in the defect with and without illumination. Therefore, we can attribute the peak f, g, h in **Fig. 4a** to the replicas of $n$ = 2, 3, 4 levels of the image potential state of the Ag(111) surface.

Our experiments clearly demonstrated the LSPR-assisted tunneling in an STM junction, providing a novel insight into a near-field induced charge transfer mechanism in a nanocavity with highly-precise control of the vacuum gap (tunneling barrier). Experiments combining STM with local optical excitation and detection also pave the way to study the near-field properties even at sub-nanometer gaps where quantum effects play a crucial role and classical electrodynamics fails to describe the near-field properties.[43, 44] The precise control of the gap distance in the STM is advantageous to study this important quantum regime of plasmon in a systematic manner. The accurate understanding of opto-electronic properties in such a "pico-cavity" is of fundamental importance in the applications for nonlinear quantum optics and optical experiments on the atomic and single-molecule level.[45] Furthermore, the image potential states can be exploited to study excited electrons dynamics.[46, 47] The fast LSPR relaxation dynamics on the femtosecond time scale is of fundamental importance to plasmon-induced hot carrier transport.[48] Ultrashort optical excitation of plasmonic STM junction[49] will provide an unprecedented opportunity to trace fast electron dynamics with high-spatial resolution.

T.K. acknowledges the support by JST-PRESTO (JPMJPR16S6). T.K. and M.W. acknowledge the support by the Deutsche Forschungsgemeinschaft through Sfb951.

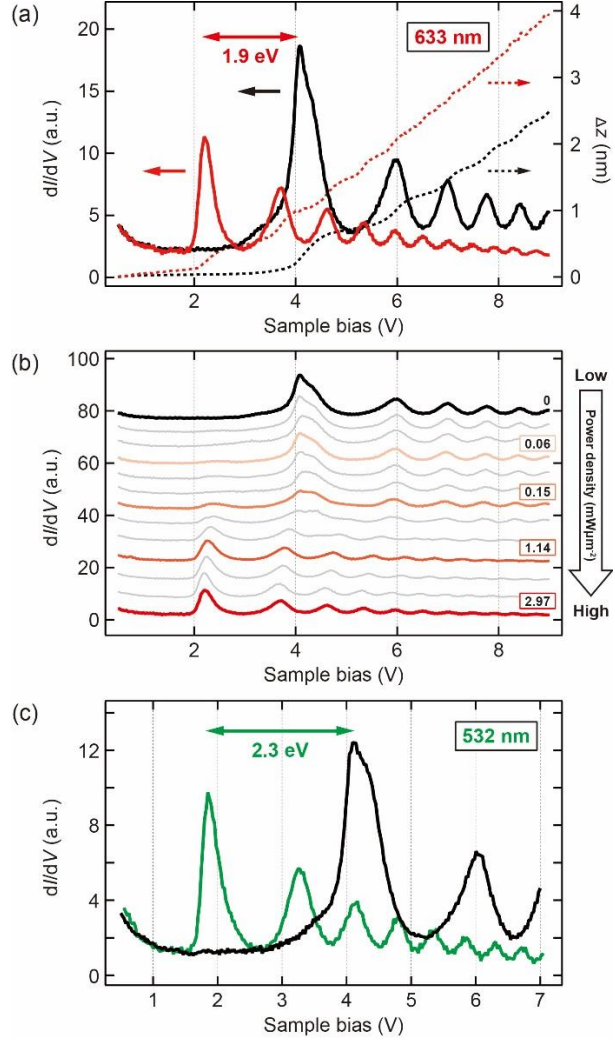

**Figure 1** (a) FER spectra (solid curves, left axis) of an Ag tip–Ag(111) surface junction with (red) and without (black) 633 nm excitation ($h\nu = 1.96$ eV). The power density is set to 2.68 mWµm$^{-2}$. The spectra were recorded in the constant current mode at 0.1 nA and the tip vertical displacements are also plotted (dashed curves, right axis). The first peak involves two components resulting from the FER and the edge of the bulk band gap.[33, 34] (b) Incident power dependence of the FER spectra. The power density is varied from 0 to 2.67 mWµm$^{-2}$. The spectra are vertically offset for clarity. (c) FER spectra (solid curves, left axis) of an Ag tip–Ag(111) surface junction with (red) and without (black) 532 nm excitation ($h\nu = 2.33$ eV). The power density is set to 5.26 mWµm$^{-2}$. The spectra were recorded in the constant current mode at 0.1 nA.

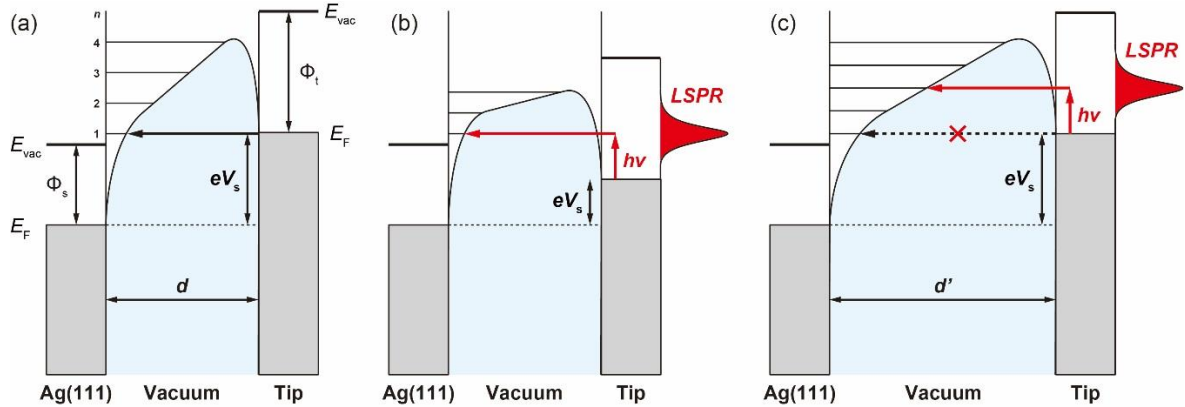

**Figure 2** Schematic model of the Ag tip–vacuum–Ag(111) junction. (a) No illumination. $E_F$: Fermi level, $E_{vac}$: vacuum level, $\Phi_{s(t)}$: work function of the surface (tip), $V_s$: Sample bias, $d$: gap distance. $n$: index of the image potential states. (b) Under illumination the LSPR assisted tunneling becomes possible. The bias voltage at which the LSPR assisted tunneling to the image potential state occurs is lower than that of (a), leading to a shallower barrier. (c) Under illumination the gap distance $d'$ is larger than $d$ in (a) when the bias voltage reaches the FER level.

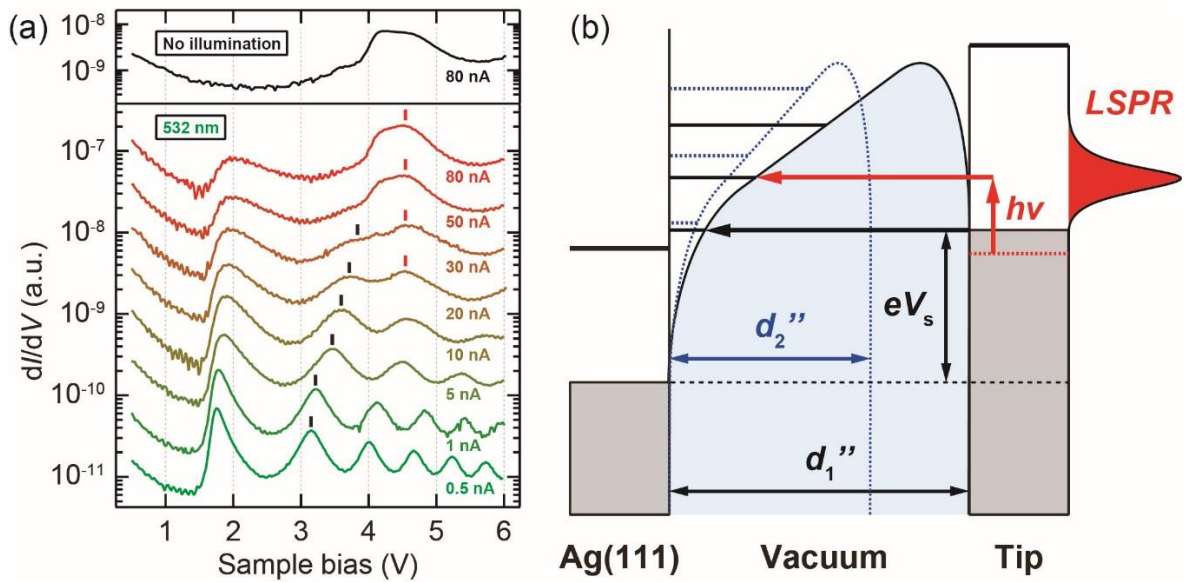

**Figure 3** (a) FER spectra of an Ag tip–Ag(111) surface junction with 532 nm excitation recorded at different gap distances (defined by the set current indicated in the figure). The power density was fixed to 5.18 mWμm$^{-2}$. The black and red bars indicate approximately the second and first peak positions of the LSPR assisted and normal tunneling process, respectively. The spectra are vertically offset for clarity. The spectrum recorded at 80 nA without illumination is also plotted in the top panel. (b) Schematic model of the Ag tip–vacuum–Ag(111) junction with a relatively small gap distance, $d_1''$ (< $d'$ in **Fig. 2c**). As the gap distance decreases ($d_2''<d_1''$), the potential becomes steeper, leading to a blue shift of FER levels as represented by the blue dashed lines. The red dashed line represents the threshold voltage at which the LSRP assisted tunneling into the second level occurs.

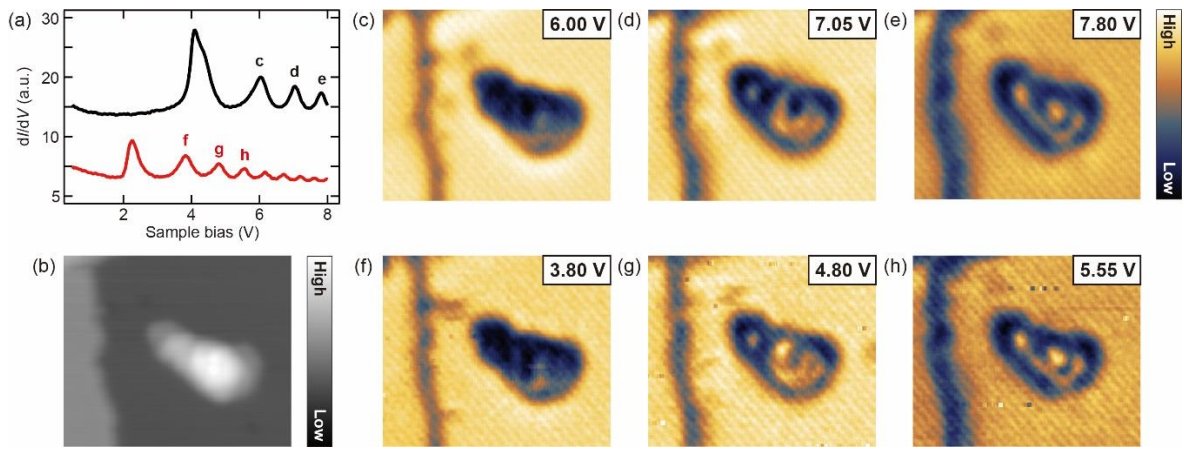

**Figure 4** (a) FER spectra of an Au tip–Ag(111) surface junction with (red) and without (black) 633 nm excitation. The spectra were recorded in the constant current mode at 0.1 nA. The power density was set to 1.02 mWµm$^{-2}$. (b) STM image of the Ag(111) surface with a monoatomic step and an intrinsic defect. (c–e) The STS mapping at different bias voltages (indicated in the figure) without illumination. (f–h) The STS mapping at different bias voltages (indicated in the figure) with illumination. The power density during the measurement was set to 1.02 mWµm$^{-2}$.

# References

[1] M. Kauranen and A. V. Zayats, Nonlinear plasmonics. *Nat. Photon.* **6**, 737-748 (2012).

[2] E. C. L. Ru and P. G. Etchegoin, Single-Molecule Surface-Enhanced Raman Spectroscopy. *Annu. Rev. Phys. Chem.* **63**, 65-87 (2012).

[3] S. Mubeen, J. Lee, W. Lee, N. Singh, G. D. Stucky, and M. Moskovits, On the Plasmonic Photovoltaic. *ACS Nano* **8**, 6066-6073 (2014).

[4] F. Tam, G. P. Goodrich, B. R. Johnson, and N. J. Halas, Plasmonic Enhancement of Molecular Fluorescence. *Nano Lett.* **7**, 496-501 (2007).

[5] Y. Tian and T. Tatsuma, Plasmon-induced photoelectrochemistry at metal nanoparticles supported on nanoporous $TiO_2$. *Chem. Commun.* 1810-1811 (2004).

[6] D. Mulugeta, K. H. Kim, K. Watanabe, D. Menzel, and H.-J. Freund, Size Effects in Thermal and Photochemistry of $(NO)_2$ on Ag Nanoparticles. *Phys. Rev. Lett.* **101**, 146103 (2008).

[7] S. Mukherjee, F. Libisch, N. Large, O. Neumann, L. V. Brown, J. Cheng, J. B. Lassiter, E. A. Carter, P. Nordlander, and N. J. Halas, Hot Electrons Do the Impossible: Plasmon-Induced Dissociation of $H_2$ on Au. *Nano Lett.* **13**, 240-247 (2013).

[8] S. Linic, U. Aslam, C. Boerigter, and M. Morabito, Photochemical transformations on plasmonic metal nanoparticles. *Nat. Mater.* **14**, 567-576 (2015).

[9] K. Ueno, T. Oshikiri, and H. Misawa, Plasmon‐induced water splitting using metallic‐nanoparticle‐loaded photocatalysts and photoelectrodes. *ChemPhysChem* **17**, 199-215 (2016).

[10] M. L. Brongersma, N. J. Halas, and P. Nordlander, Plasmon-induced hot carrier science and technology. *Nat. Nanotechnol.* **10**, 25-34 (2015).

[11] P. Christopher and M. Moskovits, Hot Charge Carrier Transmission from Plasmonic Nanostructures. *Annu. Rev. Phys. Chem.* **68**, 379-398 (2017).

[12] R. Berndt, J. K. Gimzewski, and P. Johansson, Inelastic Tunneling Excitation of Tip-Induced Plasmon Modes on Noble-Metal Surfaces. *Phys. Rev. Lett.* **67**, 3796-3799 (1991).

[13] G. V. Nazin, X. H. Qiu, and W. Ho, Atomic Engineering of Photon Emission with a Scanning Tunneling Microscope. *Phys. Rev. Lett.* **90**, 216110 (2003).

[14] A. Yu, S. Li, H. Wang, S. Chen, R. Wu, and W. Ho, Visualization of Nanoplasmonic Coupling to Molecular Orbital in Light Emission Induced by Tunneling Electrons. *Nano Lett.* DOI: 10.1021/acs.nanolett.8b00613 (2018).

[15] R. Berndt, R. Gaisch, J. K. Gimzewski, B. Reihl, R. R. Schlittler, W. D. Schneider, and M. Tschudy, Photon emission at molecular resolution induced by a scanning tunneling microscope. *Science* **262**, 1425 (1993).

<sup>16</sup> N. L. Schneider and R. Berndt, Plasmonic excitation of light emission and absorption by porphyrine molecules in a scanning tunneling microscope. *Phys. Rev.* B **86**, 035445 (2012).

[16] N. L. Schneider and R. Berndt, Plasmonic excitation of light emission and absorption by porphyrine molecules in a scanning tunneling microscope. *Phys. Rev.* B **86**, 035445 (2012).

[17] T. Lutz, C. Große, C. Dette, A. Kabakchiev, F. Schramm, M. Ruben, R. Gutzler, K. Kuhnke, U. Schlickum, and K. Kern, Molecular orbital gates for plasmon excitation. *Nano Lett.* **13**, 2846 (2013).

[18] C. Große, A. Kabakchiev, T. Lutz, R. Froidevaux, F. Schramm, M. Ruben, M. Etzkorn, U. Schlickum, K. Kuhnke, and K. Kern, Dynamic Control of Plasmon Generation by an Individual Quantum System. *Nano Lett.* **14**, 5693 (2014).

[19] X. H. Qiu, G. V. Nazin, and W. Ho, Vibrationally resolved fluorescence excited with submolecular precision. *Science* **299**, 542-546 (2003).

[20] M. C. Chong, G. Reecht, H. Bulou, A. Boeglin, F. Scheurer, F. Mathevet, and G. Schull, Narrow-Line Single-Molecule Transducer between Electronic Circuits and Surface Plasmons. *Phys. Rev. Lett.* **116**, 036802 (2016).

[21] M. C. Chong, L. Sosa-Vargas, H. Bulou, A. Boeglin, F. Scheurer, F. Mathevet, and G. Schull, Ordinary and Hot Electroluminescence from Single-Molecule Devices: Controlling the Emission Color by Chemical Engineering. *Nano Lett.* **16**, 6480-6484 (2016).

[22] Y. Zhang, Y. Luo, Y. Zhang, Y.-J. Yu, Y.-M. Kuang, L. Zhang, Q.-S. Meng, Y. Luo, J.-L. Yang, Z.-C. Dong, and J. G. Hou, Visualizing coherent intermolecular dipole–dipole coupling in real space. *Nature* **531**, 623-627 (2016).

[23] K. Kuhnke, C. Große, P. Merino, and K. Kern, Atomic-Scale Imaging and Spectroscopy of Electroluminescence at Molecular Interfaces. *Chem. Rev.* **117**, 5174-5222 (2017).

[24] H. Imada, K. Miwa, M. Imai-Imada, S. Kawahara, K. Kimura, and Y. Kim, Single molecule investigation of energy dynamics in a coupled plasmon-exciton system. *Phys. Rev. Lett.* **119**, 013901 (2017).

[25] Z. C. Dong, X. L. Zhang, H. Y. Gao, Y. Luo, C. Zhang, L. G. Chen, R. Zhang, X. Tao, Y. Zhang, J. L. Yang, and J. G. Hou, Generation of molecular hot electroluminescence by resonant nanocavity plasmons. *Nat. Photon.* **4**, 50-54 (2010).

[26] B. Pettinger, P. Schambach, C. J. Villagómez, and N. Scott, Tip-enhanced Raman spectroscopy: near-fields acting on a few molecules. *Ann. Rev. Phys. Chem.* **63**, 379-399 (2012).

[27] R. Zhang, Y. Zhang, Z. C. Dong, S. Jiang, C. Zhang, L. G. Chen, L. Zhang, Y. Liao, J. Aizpurua, Y. Luo, J. L. Yang, and J. G. Hou, Chemical mapping of a single molecule by plasmon-enhanced Raman scattering. *Nature* **498**, 82-86 (2013).

[28] E. A. Pozzi, G. Goubert, N. Chiang, N. Jiang, C. T. Chapman, M. O. McAnally, A.-I. Henry, T. Seideman, G. C. Schatz, M. C. Hersam, and R. P. Van Duyne, Ultrahigh-Vacuum Tip-Enhanced Raman Spectroscopy. *Chem. Rev.* **117**, 4961-4982 (2017).

[29] N. Tallarida, J. Lee, and V. A. Apkarian, Tip-Enhanced Raman Spectroscopy on the Angstrom Scale: Bare and CO-Terminated Ag Tips. *ACS Nano* **11**, 11393-11401 (2017).15

[30] S. W. Wu, N. Ogawa, and W. Ho, Atomic-Scale Coupling of Photons to Single-Molecule Junctions. *Science* **312**, 1362-1365 (2006).

[31] G. Binnig, K. H. Frank, H. Fuchs, N. Garcia, B. Reihl, H. Rohrer, F. Salvan, and A. R. Williams, Tunneling Spectroscopy and Inverse Photoemission: Image and Field States. *Phys. Rev. Lett.* **55**, 991-994 (1985).

[32] R. S. Becker, J. A. Golovchenko, and B. S. Swartzentruber, Electron Interferometry at Crystal Surfaces. *Phys. Rev. Lett.* **55**, 987-990 (1985).

[33] J. Martínez-Blanco and S. Fölsch, Light emission from Ag(111) driven by inelastic tunneling in the field emission regime. *J. Phys.: Condens. Matter* **27**, 255008 (2015).

[34] T. Kumagai, S. Liu, A. Shiotari, D. Baugh, S. Shaikhutdinov, and M. Wolf, Local electronic structure, work function, and line defect dynamics of ultrathin epitaxial ZnO layers on a Ag(111) surface. *J. Phys.: Condens. Matter* **28**, 494003 (2016).

[35] P. B. Johnson and R. W. Christy, Optical Constants of the Noble Metals. *Phys. Rev.* B **6**, 4370-4379 (1972).

[36] K. Giesen, F. Hage, F. J. Himpsel, H. J. Riess, and W. Steinmann, Two-photon photoemission via image-potential states. *Phys. Rev. Lett.* **55**, 300-303 (1985).

[37] P. M. Echenique and J. B. Pendry, Theory of image states at metal surfaces. *Prog. Surf. Sci.* **32**, 111-159 (1989).

[38] D. B. Dougherty, P. Maksymovych, J. Lee, M. Feng, H. Petek, and J. T. Yates, Jr. Tunneling spectroscopy of Stark-shifted image potential states on Cu and Au surfaces. *Phys. Rev.* B **76**, 125428 (2007).

[39] T. Jung, Y. W. Mo, and F. J. Himpsel, Identification of Metals in Scanning Tunneling Microscopy via Image States. *Phys. Rev. Lett.* **74**, 1641-1644 (1995).

[40] M. Pivetta, F. Patthey, M. Stengel, A. Baldereschi, and W.-D. Schneider, Local work function Moiré pattern on ultrathin ionic films: NaCl on Ag(100). *Phys. Rev.* B **72**, 115404 (2005).

[41] P. Ruffieux, K. Aït-Mansour, A. Bendounan, R. Fasel, L. Patthey, P. Gröning, and O. Gröning, Mapping the Electronic Surface Potential of Nanostructured Surfaces. *Phys. Rev. Lett.* **102**, 086807 (2009).

[42] K. Schouteden and C. Van Haesendonck, Quantum Confinement of Hot Image-Potential State Electrons. *Phys. Rev. Lett.* **103**, 266805 (2009).

[43] J. Zuloaga, E. Prodan, and P. Nordlander, Quantum Description of the Plasmon Resonances of a Nanoparticle Dimer. *Nano Lett.* **9**, 887-891 (2009).

[44] W. Zhu, R. Esteban, A. G. Borisov, J. J. Baumberg, P. Nordlander, H. J. Lezec, J. Aizpurua, and K. B. Crozier, Quantum mechanical effects in plasmonic structures with subnanometre gaps. *Nat. Commun.* **7**, 11495 (2016).

[45] F. Benz, M. K. Schmidt, A. Dreismann, R. Chikkaraddy, Y. Zhang, A. Demetriadou, C. Carnegie, H. Ohadi, B. de Nijs, R. Esteban, J. Aizpurua, and J. J. Baumberg, Single-molecule optomechanics in "picocavities". *Science* **354**, 726-729 (2016).

[46] M. Wolf, E. Knoesel, and T. Hertel, Ultrafast dynamics of electrons in image-potential states on clean and Xe-covered Cu(111). *Phys. Rev.* B **54**, R5295-R5298 (1996).

[47] U. Höfer, I. L. Shumay, Ch. Reuß, U. Thomann, W. Wallauer, and Th. Fauster, Time-Resolved Coherent Photoelectron Spectroscopy of Quantized Electronic States on Metal Surfaces. *Science* **277**, 1480-1482 (1997).

[48] S. Tan, A. Argondizzo, J. Ren, L. Liu, J. Zhao, and H. Petek, Plasmonic coupling at a metal/semiconductor interface. *Nat. Phys.* **11**, 806-812 (2017).

[49] Y. Terada, S. Yoshida, O. Takeuchi, and H. Shigekawa, Real-space imaging of transient carrier dynamics by nanoscale pump–probe microscopy. *Nat. Photon.* **4**, 869-874 (2010).